\documentclass[12pt,english]{article}
\usepackage[T1]{fontenc}
\usepackage[latin9]{inputenc}
\usepackage[paperwidth=8.5in,paperheight=11in]{geometry}
\usepackage{setspace}
\makeatletter
\usepackage{enumitem}		

\usepackage{aStyle}

\makeatother

\usepackage{babel}
\begin{document}
\begin{doublespace}
\begin{center}
\textbf{\LARGE{}Herbert Dingle and\\\vspace{0.2cm}``Science at the
Crossroads''}\vspace{-1.3cm}
\par\end{center}
\end{doublespace}

\begin{center}
Taha Sochi (Contact: ResearchGate)\vspace{-0.4cm}
\par\end{center}

\begin{center}
London, United Kingdom
\par\end{center}

\textbf{Abstract}: In this article we pay tribute to Herbert Dingle
for his early call to re-assess special relativity from philosophical
and logical perspectives.\footnote{This tribute (which was originally a collection of notes) was due
to be published in 2022 which is the jubilee of Dingle's book ``Science
at the Crossroads'' \cite{DingleBook1972}. However, there were delays
in editing and finalizing this article (due to my busy schedule in
the last two years).} However, we disagree with Dingle about a number of issues particularly
his failure to distinguish between the scientific essence of special
relativity (as represented by the experimentally-supported Lorentz
transformations and their formal implications and consequences which
we call ``the mechanics of Lorentz transformations'') and the logically
inconsistent interpretation of Einstein (which is largely based on
the philosophical and epistemological views of Poincare). We also
disagree with him about his manner and attitude which he adopted in
his campaign against special relativity although we generally agree
with him about the necessity of impartiality of the scientific community
and the scientific press towards scientific theories and opinions
as well as the necessity of total respect to the ethics of science
and the rules of moral conduct in general.

\begin{flushleft}
\par\end{flushleft}

\clearpage{}

\tableofcontents{}

\clearpage{}

\section{\label{secIntroduction}Introduction}

The controversy about the theory of special relativity dates back
to the early days of its appearance, where these controversies took
initially the form of ``paradoxes'' (e.g. the twin paradox, the
barn-pole paradox and the twirling pole paradox). Most of these controversies
(as mostly formulated and embedded in ``paradoxes'' and similar
arguments) were targeting the logical foundations and epistemological
implications of special relativity. The essence of most of these arguments
and paradoxes against the theory is the apparent contradiction between
the requirement of the principle of relativity which denies the existence
of any privileged frame of reference (in uniform relative motion)
and the requirement of certain effects and implications of the theory
to have such a privileged frame of reference.

Although the majority of scientists (representing the mainstream physics)
accepted special relativity with many of them advocating it vigorously
and enthusiastically, there were a few prominent scientists who opposed
it or expressed their discomfort and doubt about it due to the aforementioned
logical and epistemological issues. One of these opponents to special
relativity was the English physicist and philosopher of science Herbert
Dingle who dedicated his last years trying to convince the scientific
world that special relativity is logically inconsistent and hence
it is scientifically wrong (and even dangerous). His effort (which
is vividly documented in his book ``Science at the Crossroads''
\cite{DingleBook1972}) mostly took the form of communicating with
fellow scientists (either directly or indirectly through publication
of articles in scientific journals) trying to convince them of his
views and convictions which were largely based on the famous clock
(or twin) paradox.

In this article we try (within our tribute to Herbert Dingle) to highlight
the main points and issues (or concerns) of Dingle in his vigorous
campaign against special relativity where we also present our assessment
and position towards the two sides of this dispute (or debate) noting
that we agree with Dingle on certain issues and disagree with him
on other issues although we generally sympathize with his views and
campaign against special relativity (which should explain our desire
to pay tribute to him). In fact, we believe that Dingle was not fairly
treated and his views on certain issues deserve better attention and
reception from the scientific community of his time (as well as the
next generations of scientists including our present generation).

Our view about special relativity is outlined in our paper \cite{SochiSpecRel}
and discussed in detail in our book ``The Mechanics of Lorentz Transformations''
\cite{SochiBook4}. So, we have no intention to go through these details
in the present paper (i.e. we advise the interested readers to refer
to these documents for details). However, we briefly summarize our
view in the following paragraph. Our main point against both parties
(i.e. the proponents and the opponents of special relativity) is that
they both fail to distinguish between the formalism of the theory
(which is its real scientific essence represented by the Lorentz transformations
with their formal consequences and implications) and the Einsteinian
interpretation (which essentially represents the philosophical and
epistemological views of Poincare which Einstein adopted possibly
with some modifications that resulted in these logical inconsistencies).

In brief, we distinguish between the logically inconsistent theory
of special relativity as an epistemological interpretation of the
scientifically sound formalism of Lorentz transformations (which is
the authentic scientific essence of the so-called ``special relativity'').
So, while we accept the formalism of the theory (which in our view
is no more than the experimentally supported Lorentz transformations)
we reject the epistemological and philosophical content of the theory
since we regard it as an interpretation to the formalism. In this
regard, we compare special relativity to the interpretations of quantum
mechanics (such as the Copenhagen interpretation and the hidden-variable
interpretation), and compare Lorentz transformations (with all their
implications and consequences which form the formalism of the so-called
``special relativity'') to the formalism of quantum mechanics.\footnote{The duality of formalism-interpretation of scientific theories is
fully investigated and discussed (within the context of quantum mechanics)
in our book ``The Epistemology of Quantum Physics'' \cite{SochiBookQuantum}
which the interested readers should refer to.} In fact, this is the motive behind our labeling of ``special relativity''
as ``the mechanics of Lorentz transformations'' (which is the title
of our book that we referred to earlier).

\clearpage{}

\section{The Concerns of Dingle}

In the following subsections we outline the main concerns (which are
mostly presented in the form of arguments, challenges, warnings, etc.
or embedded within such forms) that Dingle presented in his campaign
against special relativity (as depicted and outlined in his book ``Science
at the Crossroads'' \cite{DingleBook1972}).

\subsection{The Clock (or Twin) Paradox}

This is the most famous of all ``paradoxes'' that were invented
and proposed to challenge the theory of special relativity. In fact,
it was proposed (and actually was the subject of discussion and controversy
in the scientific community which includes some of the most prominent
physicists of their time) since the early days of the appearance of
special relativity. This paradox is very well known and it is fully
explained and discussed in many textbooks about the relativity theories
(see for instance \cite{DauriaTBook2016,SochiBook4}) and hence we
are not going to present it or discuss it in this short paper.

However, this paradox is at the heart of Dingle's arguments against
special relativity (as presented in detail in his book) and hence
we should outline it from Dingle's perspective. The following excerpt
from Dingle's book outlines this paradox (and should actually summarize
his main challenge to special relativity):\footnote{In fact, this excerpt is just a sample representing other similar
excerpts that can be quoted to represent Dingle's argument and viewpoint.}

According to the theory, if you have two exactly similar clocks, A
and B, and one is moving with respect to the other, they must work
at different rates (a more detailed, but equally simple, statement
is given on pp. 45-6, but this gives the full essence of the matter),
i.e. one works more slowly than the other. But the theory also requires
that you cannot distinguish which clock is the `moving' one; it is
equally true to say that A rests while B moves and that B rests while
A moves. The question therefore arises: how does one determine, consistently
with the theory, which clock works the more slowly? Unless this question
is answerable, the theory unavoidably requires that A works more slowly
than B and B more slowly than A -$~\!$-which it requires no super-intelligence
to see is impossible. (End of quote)

We totally agree with Dingle about his challenge to special relativity
from this perspective. In fact, we have discussed this in detail in
our book \cite{SochiBook4}. However. this should be a challenge to
the epistemological interpretation of special relativity and not to
the formalism of Lorentz mechanics (as Dingle apparently wants). In
fact, there is nothing in the entire formalism of Lorenz mechanics
(i.e. neither in the Lorentz transformations nor in their direct or
indirect formal consequences and implications) that indicates such
a logical absurdity or allows it. So in brief, although Dingle is
completely right in his challenge to this logical and epistemological
aspect of special relativity, he should be wrong in his seeming attempt
to discredit the formalism of Lorentz mechanics (which is the scientific
essence of what is called ``special relativity''). Also see $\S$
\ref{subLorentzTrans}.

\subsection{The Role of Mathematics in Science}

We quote in this regard the following excerpt from Dingle's book (referring
the readers to his book for more quotes like this):

What I believe to be the basic misconception of modern mathematical
physicists --- evident, as I say, not only in this problem but conspicuously
so throughout the welter of wild speculations concerning cosmology
and other departments of physical science -- is the idea that everything
that is mathematically true must have a physical counterpart; and
not only so, but must have the particular physical counterpart that
happens to accord with the theory that the mathematician wishes to
advocate. (End of quote)

We generally agree with Dingle's view about the role of mathematics
in physical sciences and his opposition to the excessive mathematization
of modern science.\footnote{In fact, we disagree with Dingle about some details with regard to
this issue.} Science, after all, is an observational-experimental enterprise and
hence it should rely firstly and mostly on first hand observation
of nature and experimenting on it. Yes, the behavior of nature can
generally be mathematized through modeling and formulation but this
does not mean that nature necessarily follow our mathematical models
and formulations (at least in their entirety and with all their consequences).
So, the right course of action is to observe and experiment first,
and this should be followed by trying to put the results of our observations
and experiments in mathematical models and forms (if possible).

The danger of excessive mathematization of science is that it leads
to illusions and fantasies since mathematics is full of artifacts
and abstract objects that do not correspond to physical reality (e.g.
we have imaginary solutions, singularities, infinities, senselessly-negative
quantities, etc.). In fact, the obsession of modern scientists with
mathematics and mathematization has led to some disastrous consequences
such as wasting huge resources on investigating trivial or illusory
things and the emergence of bogus branches of science and scientific
thinking. The theories of relativity (especially general relativity)
should take part of the blame for these dangerous trends in modern
science since they are over-mathematized or require excessive mathematization
to digest some of their nonsensical consequences and implications.\footnote{We refer the readers to our books about the relativity theories (see
\cite{SochiBook4} and \cite{SochiBook5}) for more details.}

\subsection{The General Attitude of Scientific Community}

This is another concern of Dingle that we generally agree with. In
brief, there is a considerable amount of prejudice and discrimination
among the scientific community when it comes to the assessment and
treatment of scientific theories and opinions where some are favored
on the basis of non-scientific criteria and non-scholar grounds, and
this is particularly true with regard to the relativity theories which
enjoy special treatment in which the scientific and professional standards
and rules (including some ethical codes and principles) are generally
ignored or marginalized or sidelined due to the magic influence of
Einstein and the huge propaganda in favor of him and his theories
and views (which sometimes lead even to intimidation and bullying
as well as other forms of thuggish behavior).\footnote{The following extract from Dingle's book outlines this issue: The
fact that, nevertheless, a request for assistance in restoring integrity
in science can be read as a request to `try to distinguish between
the two sides' on a particular scientific point I can only regard
as one more example of the evil spell cast by the word `relativity'
--- a word that immediately reduces the mental power of even leading
physicists to impotence and is the greatest stumbling-block to my
efforts to bring home to them the extreme seriousness of the state
to which we have been reduced. Apart from that word-magic, there is
nothing in the whole course of events which I related which might
not have happened if `crystallography' had been substituted for `relativity':
it is just a historical accident that Einstein's theory caused, or
showed up, the corruption. (End of quote)\\
We also quote the following extract: Science no longer refuses to
tolerate the neglect of any anomaly; it refuses to tolerate anything
but neglect of a most outstanding anomaly. It no longer fears only
prejudice and preconception; it fears to the point of terror a particular
threat to its prejudices and preconceptions, and does everything in
its power to suppress such a threat. Its criteria of truth --- if
that word can still be used in connection with it --- are no longer
reason and experience, but strict conformity to a theory ... etc.}

In fact, the episode of Dingle in his campaign against special relativity
should highlight the dogmatic attitude of the supposedly enlightened
scientific community towards certain ``untouchable'' elements in
science (or ``holy'' beliefs) which makes discussion or opposition
to these elements a ``taboo'' and could lead to the ``excommunication''
of whoever dares to challenge them. This ``religious'' component
in modern science is one of the many disgraces and shames of modern
science and should be addressed. This dogmatic attitude is reflected
in many aspects and activities of science and scientists, e.g. in
the choice of teaching and research topics, in the determination of
what should and should not be accepted for publication in the scientific
press, in the giving of grants and awards, in the allocation of honors
and prestige, and so on. Although this unscientific attitude is not
limited to special relativity, special relativity is one of the prominent
examples of ``holy'' elements that enjoy special treatment in modern
science (since it has a ``special place'' in the heart of the mainstream
scientists).

However, we should note that the situation has improved in modern
times (compared to the time of Dingle) with the emergence of more
theories, experiments, observations, individual scientists, opinions
(... etc.) which are not in favor of special relativity and its principles
and propositions. For instance, the emergence of Bell's theorem in
quantum mechanics (and what followed of quantum entanglement experiments)
shook some of the foundations of special relativity. Also, novel experiments
and observations related to the speed of light put some question marks
on the dogmatic belief of special relativity about the speed of light
as the ultimate and invariant physical speed.

\subsection{The General Attitude of Scientific Press}

This is another concern of Dingle that we generally agree with. Scientific
press (especially the prestigious ones) generally welcomes only certain
types of papers which are supportive of the mainstream (or orthodox)
theories and opinions and rejects (generally without consideration)
any paper that advocates or sympathizes with opposite (or non-orthodox)
theories and opinions, and this is particularly true when it comes
to the relativity theories where any paper that questions these theories
or challenges them is usually (and automatically) rejected as a form
of heresy or pseudo-science or crackpot idiocy.

The obvious result is that only certain views and trends in science
are heard and sponsored and hence only certain views and trends are
encouraged to grow and prosper since the opposite views and trends
will naturally die or weaken because very few scientists have the
determination to continue their work when this work has no chance
(or very little chance) to be in the public space. This form of unnatural
selection is not only against the spirit and ethics of science but
it is also against the interest of science and humanity since valuable
ideas and contributions can be lost as a result of this form of unfair
discrimination.

However, the situation generally improved in the recent times thanks
to the emergence and wide availability of pre-print repositories and
social networks (professional as well as non-professional) which enable
anyone to publish his work. However, the attention and seriousness
that published work enjoys when it is published in the mainstream
and prestigious media (e.g. the high impact journals) still favors
views and opinions that represent and accept (or inline with) the
mainstream science, and this is certainly a type of prejudice and
favoritism that should be addressed and fixed.

As indicated earlier, this discriminatory attitude is not limited
to the scientific press but it can be seen and felt in every department
of modern science such as awards, grants, academic posts, scientific
venues (like conferences) and so on. In fact, the attitude of scientific
press is just one of many forms and examples of discrimination and
unfair treatment that characterize all the departments of modern science
(noting that we limited our attention to the attitude of scientific
press because that is what concerned Dingle in his book).

\subsection{The Ethics of Modern Science}

One of the most important (and ``most true'') concerns of Dingle
(which he documented in detail in his book, sometimes explicitly and
directly and sometimes implicitly and indirectly) is the lack of transparency
and the absence of respect to the rules of ethics and morality in
many practices of modern scientists. Unfortunately, science in modern
times (and possibly even in old times) seems to be no more than a
way for making a living and getting financial gains, reputation, prestige,
awards, and so on. So, the search for truth and service to humanity
(which are supposed to be, or at least should be, the main purpose
of science) are of no concern to many (if not most) scientists.

In my view, the dodgy practices and unethical behavior of scientists
of the kind indicated and highlighted by Dingle is just one type of
the many types of immorality (and possibly the slightest type of immorality)
of modern science and scientists. In fact, modern science and scientists
are embroiled in many other types of immorality (and even criminality)
such as experimenting on animals (or rather torturing and killing
them in the name of science) and developing lethal weapons (especially
weapons of mass destruction and weapons designed to target civilians
or necessarily damage civilians and civilian infrastructures). So,
despite the alleged ethical codes and rules of morality which are
supposedly adopted by ``respected'' scientific institutions and
bodies, the reality is very different especially when we note that
even if all these alleged codes of conduct are implemented literally
they do not satisfy even the minimum requirements of real ethical
conduct and moral behavior because immorality and criminality are
at the very essence of some of these ``scientific'' practices and
disciplines. Anyway, this is not our main concern in this paper (but
we intend to come back to this issue in the future).

\subsection{\label{subLorentzTrans}The Transformations of Lorentz}

Dingle seems to reject not only special relativity as an interpretation
(demonstrated for instance in some of its epistemological aspects
and consequences) as we do, but he rejects even its formalism. Although
we do not believe that the formalism of Lorenz mechanics (which is
based on the transformations of Lorentz as postulates in our view)
is a final theory and it may not be a complete theory, we believe
that the accumulated observational and experimental evidence should
qualify this formalism to be regarded as an accepted scientific theory
with the rejection of the illogical epistemological interpretation
that is attached to it (as represented by special relativity).

In fact, Dingle seems to put question marks even on Maxwell's electrodynamic
theory which proved to be very successful over its long history (in
theory as well as in application and practice). This view of Dingle
is understandable when we note the intimate relationship between Lorentz
transformations and Maxwell's electrodynamic theory. Although it is
possible and acceptable (in principle) to replace Maxwell's electrodynamic
theory within a ``global'' and radical change (or rather revolution)
in modern physics, it seems unwise (as well as impractical) to target
Maxwell's theory initially and in this manner because of a logical
inconsistency in special relativity. In other words, we can easily
dismiss special relativity and get rid of it without sacrificing the
formalism of Maxwell's equations or the Lorentz transformations, and
this should not only be the easiest (or even the very easy) thing
to do but also the most wise option. In fact, it is the option that
is most consistent with the spirit and tradition of science and how
it should develop and progress.

\subsection{The Danger of Wrong Theory}

The following excerpt from Dingle's book should summarize one of his
major concerns about special relativity:

The second reason for the publication of this book is a practical
one. Directly or indirectly - at present chiefly the latter, though
none the less inseparably - special relativity is involved in all
modern physical experiments, and these are known to be attended by
such dangerous possibilities, should something go wrong with them,
that the duty of ensuring as far as possible that this shall not happen
is imperative. It is certain that, sooner or later, experiments based
on false theories will have unexpected results, and these, in the
experiments of the present day, may be harmless or incalculably disastrous.
In these circumstances an inescapable obligation is laid on experimental
physicists to subject their theories to the most stringent criticism.
As this book will show, their general practice is to leave such criticism
to mathematical theorists who either evade or ignore it, and the possible
consequences are evident and unspeakably menacing. This alone would
compel the publication of the facts here revealed. (End of quote)\footnote{We should refer the readers to Dingle's book \cite{DingleBook1972}
for more quotes in this regard.}

In fact, this is one of the strangest claims of Dingle because if
``special relativity'' is a scientifically correct theory then it
should already be ``tested'' in the laboratory of Universe and hence
any new (human-made) experiment based on it should not lead to such
disastrous consequences (as Dingle claims) more than any other experiment
(based on other correct theory) do. On the other hand, if ``special
relativity'' is a scientifically incorrect theory then any danger
from experiments based on it should be an illusion because the theory
should not lead to the consequences that the theory is supposed to
produce. So, any danger from such experiments should originate (if
so) from other sources of danger which all scientific experiments
share (i.e. any scientific experiment can go wrong because of its
bad design or because of its bad theoretical formulation or because
of many other things like these). In fact, someone may argue exactly
the opposite by claiming that false theories should be safer than
correct theories because the expected effects of false theories should
be null (or void).

\clearpage{}

\section{The Approach and Attitude of Dingle Himself}

Although we appreciate the enthusiasm of Dingle in his campaign against
special relativity and his attempt to convince and persuade the scientific
world of its falsehood and ``danger'', we generally disagree with
him about the approach that he followed and the attitude that he adopted
in this campaign. We note for example that he personalized the dispute
in some occasions and gave it the taste and flavor of a fight instead
of keeping it as a theoretical scientific debate. Moreover, he was
relentlessly insistent and repetitive in his demands trying to force
his views and convictions and impose his wishes on other people who
disagree with him or do not share his views and convictions.

However, we should forgive and excuse Dingle in this regard considering
that special relativity in his view was not only a theoretical or
logical ``danger'' to science but it is a physical danger to humanity
and its physical safety and existence. The factor of age should also
be considered and respected since a man of his age (at that time)
can easily lose patience and temper and become somewhat detached from
reality. In my view, the episode of Dingle in his campaign against
special relativity should be seen as an amusing and entertaining story
that should add a touch of humor and humanity on ``cold'' and ``emotionless''
science instead of being treated as an example of naughtiness and
misbehavior and remembered with bitterness and discomfort (as reflected
in the attitude of some opponents of Dingle whether those who were
engaged in that dispute or those who expressed their opinion about
it later on).

Anyway, Dingle's book contains a considerable amount of nonsense and
useless arguments. It was better for him to present his views about
special relativity (as well as about the integrity of science, scientists,
scientific press, the ethics of science and so on) in a more professional
way and academic manner instead of this talkative and vulgar way.

\clearpage{}

\section{Conclusions}

We summarize the main issues discussed or implicated in the present
article in the following bullet points:

\noindent $\bullet$ In our view, ``special relativity'' (as it
is commonly known) is a combination of the experimentally-endorsed
formalism of Lorentz transformations and the logically-inconsistent
interpretation of Einstein (which originally belongs to Poincare with
the possibility of Einstein introducing some modifications or elaborations
on it).\footnote{In fact, Einstein's version of the theory may not represent the views
of Poincare exactly (since we have no sufficient knowledge about some
details of Poincare's views). So, it is possible that the views of
Poincare may not be logically inconsistent or they may not share the
same logical inconsistencies as those in the interpretation of Einstein.
In other words Einstein's version of ``special relativity'' may
not be exactly the same as Poincare's version of ``special relativity''.} So, while we criticize and reject the logically-inconsistent interpretation
of Einstein we think that the formalism of the theory (which we call
``the mechanics of Lorentz transformations'') is generally acceptable
theory (although it is possibly a limited and interim theory).

\noindent $\bullet$ In our view, the main failure in Dingle's thesis
is his failure to distinguish between the formalism of special relativity
(which essentially is no more than Lorentz transformations with their
implications and consequences) and its interpretation as an epistemological
paradigm whose main purpose is to make sense of the formalism. In
other words, Dingle was unable to distinguish between the scientific
essence of the theory of relativity (i.e. the formalism of Lorentz
transformations) which is experimentally supported and the philosophical
(or epistemological) interpretation of Einstein which is logically
inconsistent.

\noindent $\bullet$ Despite Dingle's failure to distinguish between
the logically inconsistent interpretation of special relativity and
the scientifically sound formalism of Lorentz transformations (which
is the real scientific essence of what is commonly called ``special
relativity''), he should still get the credit for highlighting some
important issues about the theory of special relativity, and for this
reason we pay our tribute to him in this article. Interestingly and
fortunately, there is more acceptance to the challenges to special
relativity these days than fifty years ago and Dingle has obviously
contributed positively to this shift in the attitude towards special
relativity.

\noindent $\bullet$ Dingle's work and activities (which are mostly
documented or indicated in his book ``Science at the Crossroads'')
related to his opposition to special relativity highlight a very important
issue, that is the resistance (with stubbornness) of the ``mainstream''
science and scientists to listen to arguments and consider ideas and
theories that run outside their ``stream'' or against it. In fact,
non-scientific (and even nasty) methods and tactics are used (shamelessly)
by some scientists to suppress such opposition voices and this is
especially true when it comes to the theories of relativity where
Einstein is commonly regarded as a ``holy'' figure or symbol of
science.

\noindent $\bullet$ We generally agree with Dingle about several
other issues which include: the excessive mathematization of science,
the general attitude of scientific press (or rather the general attitude
of almost all institutions and departments of modern science), and
the unethical practices in science.

\noindent $\bullet$ We reject the fears of Dingle about potential
disaster happening as a result of accepting special relativity and
experimenting on it. We also reject the depiction of his campaign
against special relativity as a campaign to protect the public and
save humanity. Most of his fears were baseless (as particularly related
to special relativity) although warning of potential dangers of any
scientific theory of serious experimental impacts and consequences
is generally good.

\noindent $\bullet$ We also disagree with Dingle on many of his detailed
accounts and analyses. However, we did not go through these due to
the limits on the size of the paper and their relative insignificance
(e.g. because they are out of scope or because of their similarity
with the issues that we already discussed or indicated).

\noindent $\bullet$ There seems to be some confusion among the experts
on relativity (including some of those who communicated and engaged
with Dingle) between classical physics and \textit{classical} logic
where the rejection (or modification) of some classical physical concepts
(like Newtonian time and space) seems to justify to those who reject
these concepts to reject or sideline the rules of logic which should
be universally and unequivocally accepted and respected to avoid nonsensical
thinking and argumentation.

\pagebreak{}

\phantomsection 
\addcontentsline{toc}{section}{References}\bibliographystyle{plain}
\bibliography{Bibl}

\end{document}